\documentclass{emulateapj}

\shorttitle{Physical properties of Lyman-alpha emitters at $z\sim 0.3$ from UV to FIR measurements.}
\shortauthors{Oteo et al.}

\begin{document}

\title{Physical properties of Lyman-alpha emitters at $z\sim 0.3$ from UV-to-FIR measurements.}

\author{I. Oteo\altaffilmark{1,2}, 
A. Bongiovanni\altaffilmark{1,2}, 
A. M. P\'erez Garc\'{\i}a\altaffilmark{1,2}, 
J. Cepa\altaffilmark{2,1}, 
A. Ederoclite\altaffilmark{12}, 
M. S\'anchez-Portal\altaffilmark{3}, 
I. Pintos-Castro\altaffilmark{1,2}, 
R. P\'erez-Mart\'inez\altaffilmark{4}, 
D. Lutz\altaffilmark{5}, 
B. Altieri\altaffilmark{3}, 
P. Andreani\altaffilmark{8}, 
H. Aussel\altaffilmark{9}, 
S. Berta\altaffilmark{5}, 
A. Cimatti\altaffilmark{10}, 
E. Daddi\altaffilmark{9}, 
D. Elbaz\altaffilmark{9}, 
N. F\"orster Schreiber\altaffilmark{5}, 
R. Genzel\altaffilmark{5}, 
E. Le Floc'h\altaffilmark{9}, 
B. Magnelli\altaffilmark{5}, 
R. Maiolino\altaffilmark{11}, 
A. Poglitsch\altaffilmark{5},
P. Popesso\altaffilmark{5}, 
F. Pozzi\altaffilmark{6}, 
L. Riguccini\altaffilmark{7}, 
E. Sturm\altaffilmark{5}, 
L. Tacconi\altaffilmark{5}, 
I. Valtchanov\altaffilmark{3}
}

\altaffiltext{1}{Instituto de Astrof{\'i}sica de Canarias (IAC), E-38200 La Laguna, Tenerife, Spain}
\altaffiltext{2}{Departamento de Astrof{\'i}sica, Universidad de La Laguna (ULL), E-38205 La Laguna, Tenerife, Spain}
\altaffiltext{3}{Herschel Science Centre (ESAC). Villafranca del Castillo, Spain}
\altaffiltext{4}{XMM/Newton Science Operations Centre (ESAC). Villafranca del Castillo. Spain.}
\altaffiltext{5}{Max-Planck-Institut f\"{u}r Extraterrestrische Physik (MPE), Postfach 1312, 85741 Garching, Germany}
\altaffiltext{6}{INAF - Osservatorio Astronomico di Bologna, via Ranzani 1, I-40127 Bologna, Italy}
\altaffiltext{7}{Laboratoire AIM, CEA/DSM-CNRS-Universit{\'e} Paris Diderot, IRFU/Service d'Astrophysique, B\^at.709, CEA-Saclay, 91191 Gif-sur-Yvette Cedex, France}
\altaffiltext{8}{ESO, Karl-Schwarzchild-Str. 2, D-85748 Garching, Germany}
\altaffiltext{9}{Commissariat \`a l'\'Energie Atomique (CEA-SAp) Saclay, France}
\altaffiltext{10}{Dipartimento di Astronomia, Universit\`a di Bologna, Via Ranzani 1, 40127 Bologna, Italy}
\altaffiltext{11}{INAF - Osservatorio Astronomico di Roma, via di Frascati 33, 00040 Monte Porzio Catone, Italy}
\altaffiltext{12}{Centro de Estudios de F'sica del Cosmos de Arag—n, Plaza San Juan 1, Planta 2, Teruel, 44001, Spain}
 
\begin{abstract}
The analysis of the physical properties of low-redshift Ly$\alpha$ emitters (LAEs) can provide clues in the study of their high-redshift analogues. At $z \sim 0.3$, LAEs are bright enough to be detected over almost  the entire electromagnetic spectrum and it is possible to carry out a more precise and complete study than at higher redshifts. In this study, we examine the UV and IR emission, dust attenuation, SFR and morphology of a sample of 23 GALEX-discovered star-forming (SF) LAEs at $z \sim 0.3$ with direct  UV (GALEX), optical (ACS) and FIR (PACS and MIPS) data. Using the same UV and IR limiting luminosities, we find that LAEs at $z\sim 0.3$ tend to be less dusty, have slightly higher total SFRs, have bluer UV continuum slopes, and are much smaller than other galaxies that do not exhibit  Ly$\alpha$ emission in their spectrum (non-LAEs). These results suggest that at $z \sim 0.3$ Ly$\alpha$ photons tend to escape from small galaxies with low dust attenuation. Regarding their morphology, LAEs belong to Irr/merger classes, unlike non-LAEs. Size and morphology represent the most noticeable difference between LAEs and non-LAEs at $z \sim 0.3$. Furthermore, the comparison of our results with those obtained at higher redshifts indicates that either the Ly$\alpha$ technique picks up different kind of galaxies at different redshifts or that the physical properties of LAEs are evolving with redshift.
\end{abstract}


\section{Introduction}

In recent years, a substantial number of Ly$\alpha$-emitting galaxies (LAEs) have been discovered over a wide range of redshifts, from the local Universe up to $z \sim 7$, and even beyond. At $z \gtrsim 2.0$, the Ly$\alpha$ emission is located in the optical or near-IR and LAEs are mostly found using the narrow-band technique, where a combination of narrow- and broad-band filters are employed to sample the Ly$\alpha$ emission and constrain its nearby continuum, respectively \citep[e.g.,][]{Guaita2010,Bongiovanni2010,Ouchi2008,Cowiehu1998, Gawiser2006, Gronwall2007, Guaita2010, Shioya2009, Murayama2007,Ouchi2010,Nakamura2011,Sobral2009}. At $z\lesssim 2$, the Ly$\alpha$ line is in the UV and, so far, LAEs can only be found via \emph{Galaxy Evolution Explorer} \citep[GALEX,][]{Martin2005} grism spectroscopy or other space-borne UV observatories. In this case, the selection technique is based on searching for Ly$\alpha$ emission in the UV spectrum of objects with a measured UV continuum \citep{Deharveng2008,Cowie2010,Cowie2011}.

The physical properties of star-forming (SF) LAEs have been studied by classically  fitting \cite{Bruzual2003} (hereafter BC03) templates to their observed spectral energy distribution (SED) \citep[e.g.,][]{Ono2010,Lai2008,Finkelstein2008,Nilsson2007,Lai2007,Pirzkal2007,Nilsson2009,Nilsson2011,Guaita2011}. With this method, the stellar mass and age of galaxies can be reasonably constrained with  good sampling of the rest-frame UV to mid-IR SEDs, but other physical properties, such as dust attenuation, star formation rate (SFR), and metallicity tend to suffer from large uncertainties: metallicities can be obtained only by using rest-frame optical emission lines, and information about the dust emission in the FIR is essential for obtaining accurate values of dust attenuation and SFR.

At $z \sim 0.3$, \cite{Cowie2010}, \cite{Finkelstein2011_espectros} and \cite{Cowie2011} agree in their optical spectroscopic studies that LAEs are metal-poor galaxies with low dust attenuation. \cite{Cowie2011} also report that LAEs with higher EW(H$\alpha$) have bluer colors, lower metallicities and less extinctions, consistent with their being at a primeval evolutionary stage. Employing a SED- fitting procedure, \cite{Cowie2011} and \cite{Finkelstein2011_espectros} report that LAEs are mainly young galaxies with median values of 100 and 300 Myr, respectively.

With the launch of the \emph{Herschel Space Telescope} \citep{Pilbratt2010} and the data taken with the Photodetector Array Camera and Spectrometer \citep[PACS,][]{Poglitsch2010}, we are in possession of deep FIR data that allow us to constrain the FIR SED of LAEs. Unfortunately, few FIR counterparts for LAEs at $z \gtrsim 2.0$ have been reported so far; therefore, a deep analysis has not yet been possible. In fact, in \cite{Oteo2010_z2} we only found four SF LAEs at $2.0 \lesssim z \lesssim 3.5$ with FIR counterparts out of a sample of 140, only two of them having a Ly$\alpha$ rest-frame equivalent width (Ly$\alpha$ EW$_{rest-frame}$) above 20\AA, the typical minimum value for LAEs found via narrow-band imaging. Despite this low number, the detection of some LAEs in the FIR reveals that some of them are red and dusty objects, thus dust and Ly$\alpha$ emission are not mutually exclusive \citep[see also][]{Chapman2005}.

At $z \sim 0.3$, the typical FIR-observed fluxes of LAEs make them probably detectable under the limiting fluxes of PACS and MIPS--24$\mu$m observations. In \cite{Oteo2011_letter}, we report PACS-FIR detections and study dust attenuation in a sample of twelve spectroscopically GALEX-selected LAEs at $z \sim 0.3$ and $\sim 1.0$. In this paper, we expand the work started in \cite{Oteo2011_letter} about the physical properties of LAEs at z$\sim$0.3. Specifically, we obtain UV and total IR luminosities, dust attenuation, SFR, size and examine the morphology of 23 IR-detected LAEs. We also compare the results with those obtained for a sample of galaxies which, under the same limiting UV and IR luminosities, do not show  Ly$\alpha$ emission in their rest-frame UV spectra.

The structure of this paper is as follows. In Section \ref{sample} we present the LAE data sample, reporting their PACS and MIPS--24$\mu$m counterparts in Section \ref{counterparts}. The comparison sample is presented in Section \ref{control_sample}. Section \ref{results} gives the results of our study and in Section \ref{conclu} we show the main conclusions of this work. 

Throughout this paper we assume a flat universe with $(\Omega_m, \Omega_\Lambda, h_0)=(0.3, 0.7, 0.7)$, and all magnitudes are listed in the AB system \citep{Oke1983}.

\section{LAE sample}\label{sample}

In this study, we use a sample of GALEX-discovered LAEs at $z \sim 0.3$ built in \cite{Cowie2010} by looking for an emission line in the FUV spectra of objects with a measured UV continuum. Since the FUV spectra become very noisy at the edges of the spectral range, only those objects with  Ly$\alpha$ emission within 1452.5--1750 \AA\ were selected. In the redshift space, this wavelength range implies that LAEs are distributed within $z=0.195$--0.44, with a median value of $z \sim 0.3$. In \cite{Cowie2010}, LAEs were classified as SF or AGN via rest-frame UV emission-line diagnostics (shape and width of the Ly$\alpha$ line and the presence/absence of AGN ionization lines) and their X-ray measurements. In this study, we consider only those LAEs with an SF nature, ruling out the AGN. Among the fields studied in \cite{Cowie2010} we focus on COSMOS, ECDF-South, ELAIS-S1, and Lockman. These fields were selected by their wealth of FIR data from PACS-\emph{Herschel} or MIPS-\emph{Spitzer} (see Section \ref{counterparts} for more details). 

With the aim of comparing with narrow-band selected high-redshift LAEs, we only include in the sample those Ly$\alpha$-emitting galaxies whose Ly$\alpha$ EW$_{\rm rest-frame}$ are above 20 \AA, which is the typical threshold in narrow-band searches. In this sense, we define LAEs as those Ly$\alpha$-emitting galaxies whose Ly$\alpha$ EW$_{\rm rest-frame}$ are above 20 \AA, using this definition throughout the study. All the previous considerations render a sample of 30 SF LAEs. This sample is nearly complete up to $m_{\rm NUV} \sim 21.5$ mag, which represents the UV limiting magnitude of the sample. This implies a UV limiting magnitude of $\log(L_{\rm UV}/L_\odot) = 9.9$ at $z \sim 0.3$, which is similar to that in the samples of high-redshift LAEs found via the narrow-band technique.

All our LAEs  have Ly$\alpha$ luminosities below 10$^{43}$ erg s$^{-1}$, the typical median value for LAEs at $z \gtrsim 2.0$. Indeed, the lower values of the Ly$\alpha$ luminosities for LAEs at $z \sim 0.3$ are an indication of an evolution in the physical properties of LAEs between that redshift and $z\gtrsim 2.0$ \citep[see][]{Cowie2011}.

\section{PACS-FIR and MIPS--24$\mu$m counterparts of LAEs}\label{counterparts}

      \begin{figure}
   \centering
   \includegraphics[width=0.47\textwidth]{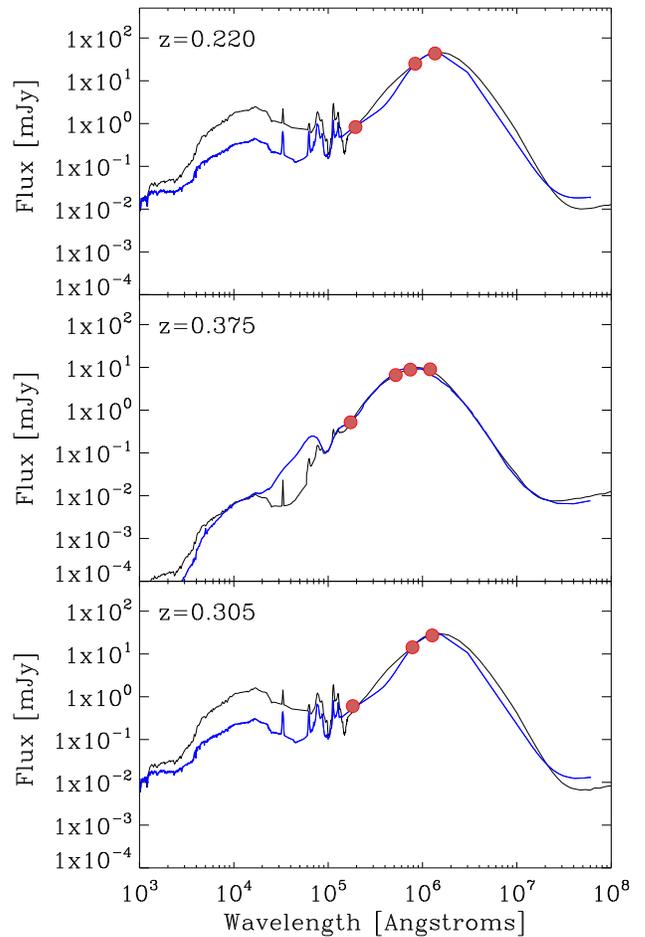}
   \caption{Observed mid-IR/FIR SEDs of the three star-forming LAEs with at least three measurements in MIPS/PACS. Red dots are MIPS/PACS observed fluxes. The black and blue curves are the \cite{Chary2001} and \cite{Polletta2007} templates, respectively, which best fit the observed photometry. The redshift of each LAE is also shown.
              }
         \label{t_dust}
   \end{figure}

The PACS-FIR observations used in this study were taken within the framework of PACS Evolutionary Probe project (PEP, PI D. Lutz). PEP is the \emph{Herschel} Guaranteed Time Key-Project to obtain the best profit from studying FIR galaxy evolution with \emph{Herschel} instrumentation \citep{Lutz2011}. Among the fields where LAEs catalogued in \cite{Cowie2010} are located, COSMOS, GOODS-South and ECDF-South have been already observed in PACS--100$\mu$m and PACS--160$\mu$m bands\footnote{The Lockman field has also been observed with PACS, but the only SF LAE at z$\sim$0.3 catalogued in \cite{Cowie2010} in that field has no detection in any of its bands.}. In this study, we use PACS fluxes extracted by using MIPS--24$\mu$m priors, with limiting (100$\mu$m, 160$\mu$m) fluxes in COSMOS, GOODS-South and ECDF-South (5.0, 11.0) mJy, (1.1, 2.0) mJy and (4.2, 8.2) mJy, respectively. The PACS-FIR counterparts of the LAEs at $z \sim 0.3$ in COSMOS and GOODS-South have already been reported in \cite{Oteo2011_letter}, by looking for a PACS detection within 2$''$ around the location of each LAE, which is the typical astrometric uncertainty in the position of the FIR  sources. In this study, we add the PACS counterparts in ECDF-South to the previous set,  following the same matching criterion. We find one extra LAE detected in both PACS--100$\mu$m and PACS--160$\mu$m, assembling, in total, a sample of six PACS-detected SF LAEs. All these FIR counterparts represent a direct measurement of the FIR dust emission in LAEs at $z \sim 0.3$. As an example, we show in Figure \ref{t_dust} the FIR SED of the three LAEs with counterparts both in PACS--100$\mu$m and PACS--160$\mu$m. Both observational FIR data points and best \cite{Chary2001} and \cite{Polletta2007} fitted templates are represented. 

MIPS--24$\mu$m measurements allow us to constrain the FIR SED of  PACS-undetected but MIPS--24$\mu$m detected LAEs. In \cite{Oteo2011_letter} we report MIPS--24$\mu$m counterparts for LAEs in the COSMOS field by using data taken from the S--COSMOS survey \citep{Sanders2007}. We found that 90\% of LAEs at $z \sim 0.3$ were detected in that band. Here, we extend the MIPS--24$\mu$m counterparts with data coming from the Spitzer Wide-area InfraRed Extragalactic survey \citep[SWIRE,][]{Lonsdale2003} in the ECDF-South, Lockman and ELAIS-South fields, using a matching radius of 2$''$ as well. We find thirteen extra MIPS-detections.

\begin{figure}
   \centering
   \includegraphics[width=0.47\textwidth]{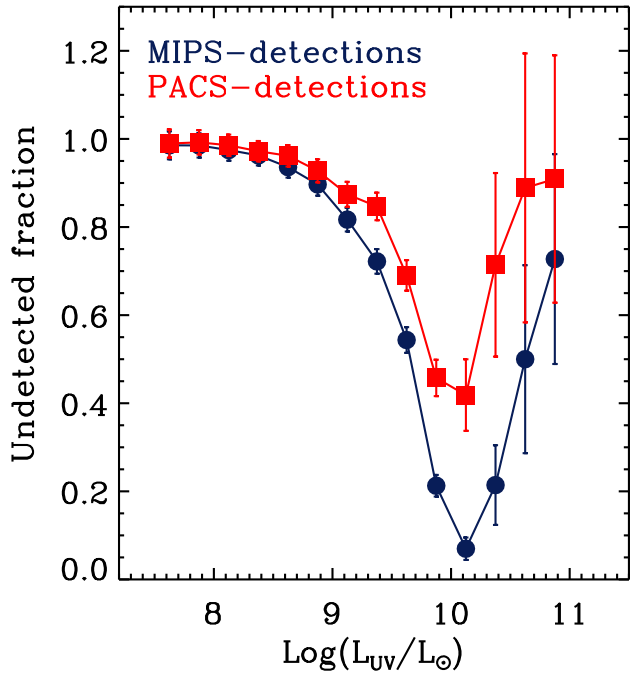}
   \caption{MIPS--24$\mu$m (blue dots) and PACS (red squares) non-detection fractions as a function of UV luminosity for a sample of UV-selected galaxies at $z\sim 0.3$. Error bars are obtained by assuming  Poisson statistics.
              }
         \label{mipsdetections}
   \end{figure}

Summarizing, among the 30 LAEs with MIPS/PACS coverage, 23 are detected in MIPS/PACS. Nine of them belong to the sample in \cite{Oteo2011_letter}. Therefore, we find that more than 75\% of LAEs at $z \sim 0.3$ are detected in the mid--IR/FIR. Figure \ref{mipsdetections} shows the detection fraction in MIPS--24$\mu$m and PACS bands of a sample of UV-selected galaxies at $z \sim 0.3$ (see Section \ref{control_sample}). It can clearly be seen that UV-brighter galaxies are more likely to be detected in the IR. This tendency is similar to that found for MIPS--24$\mu$m detections in high-redshfift galaxies \citep{Reddy2010}. There is a decrease in the detection fraction at the highest UV luminosities but in those cases the number of galaxies in each luminosity bin is low and the error bars are too large to draw any conclusion. As commented above, the limiting UV luminosity associated with the limiting magnitude of the LAEs in the sample studied is about 9.9 at $z \sim 0.3$. From Figure \ref{mipsdetections} it can be deduced that UV luminosities above that value correspond to a high detection fraction in the IR, which is compatible with the large percentage of IR-detected LAEs found in this study.

According to the limiting fluxes of the MIPS--24$\mu$m observations, MIPS-detected LAEs at $z \sim 0.3$ have $\log(L_{\rm IR}/L_\odot) \gtrsim 10.0$ in COSMOS and $\gtrsim 10.40$ in the SWIRE fields. Regarding PACS observations, the limiting luminosities are $\log(L_{\rm IR}/L_{\odot}) \sim 10.5$, $\sim 9.76$ and $\sim 10.4$ in COSMOS, GOODS-South and ECDF-South, respectively. In this sense, we adopt a limiting luminosity of $\log(L_{\rm IR}/L_{\odot}) \sim 10.4$ for the whole sample. Note that, although PACS observations in GOODS-South and MIPS observations in COSMOS are deeper than that threshold, there is only one LAE with $L_{\rm IR}$ below $10^{10.4} L_\odot$ and it is due to its lower redshift, $z \sim 0.2$. The adopted IR limiting luminosity represents an IR star formation rate of SFR$_{\rm IR} \sim 4.5 M_\odot \textrm{yr}^{-1}$, according to the \cite{Kennicutt1998} calibration.

\section{The control sample}\label{control_sample}

We also aim at comparing the derived physical properties of our IR-detected SF LAEs with those of other IR-detected SF galaxies in the same redshift range and with the same UV and IR limiting luminosities but which do not exhibit  Ly$\alpha$ emission in their UV spectrum. To do that, we define a \emph{control sample} focusing on the COSMOS field and using data from GALEX \citep{Guillaume2006,Zamojski2007} and PACS observations plus the photometric redshifts, $z_{\rm phot}$, of the COSMOS photometric catalog \citep{Capak2007}. At $z \sim 0.3$, $z_{\rm phot}$ are quite reliable and can be used instead of spectroscopic surveys, which contain much fewer objects. In this way, we select all the sources with $z_{\rm phot}$ between 0.2 and 0.4 with measurements both in GALEX and PACS and whose UV luminosities are higher than the UV limiting luminosity of LAEs at $z \sim 0.3$. In order to avoid  possible contamination from AGN, we rule out from the sample those samples which are detected in X-rays with CHANDRA. From now on, we refer to those galaxies as \emph{non-LAEs}, and they will be the main source of comparison to study the differentiating characteristic of LAEs. The sample contains 135 galaxies. This sample also contains Ly$\alpha$-emitting galaxies whose Ly$\alpha$ EW$_{\rm rest-frame}$ are below 20 \AA ; therefore, although they exhibit a Ly$\alpha$ line, it is not strong enough to be selected in narrow-band searches.

Furthermore, we also retain those galaxies with z$_{\rm phot}$ between 0.2 and 0.4, which are detected in GALEX and PACS but have UV luminosities fainter than the UV limiting luminosity of LAEs. We call these \emph{UV-faint SF galaxies}. These sources will not be directly compared with LAEs because of their different selection criterion in the UV, but will be used to place LAEs within a more general scenario of SF galaxies at the same redshift. Note that, owing to GALEX spectroscopic limitations, we do not have UV spectra for these sources and we therefore do no know whether they exhibit Ly$\alpha$ emission.

It should be noted that, for the same reasons as pointed out in Section \ref{counterparts}, there is no bias in the IR selection between LAEs and the galaxies in the control sample, all of them being  limited to the same IR luminosity, $\log(L_{\rm IR}/L_{\odot}) \sim 10.4$.

To summarize, we have a control sample formed by two kinds of galaxies: i) \emph{non-LAEs}, which are galaxies that do not show Ly$\alpha$ emission with Ly$\alpha$ EW$_{\rm rest-frame}$ above 20 \AA, are in the same redshift range of our LAEs, and have the same UV and IR limiting luminosities; ii) \emph{UV-faint SF galaxies}: formed by galaxies which are at the same redshift as our LAEs, have the same IR limiting luminosity, but are fainter in the UV than our LAEs and non-LAEs. Only the non-LAEs will be used directly to compare the properties of LAEs with those of other galaxies without Ly$\alpha$ emission. The UV-faint galaxies will be used only to place LAEs and non-LAEs within a more general framework of SF galaxies at $z \sim 0.3$.

\section{Results}\label{results}

The UV and mid-IR/FIR measurements, which represent the emitted light from young populations and the re-emission of the light absorbed by dust in the UV, respectively, can be used to analyze the physical properties of our galaxies. The UV/IR combination allows an accurate determination of dust attenuation and total SFR, SFR$_{\rm total}$. Furthermore, FIR detections themselves enable us to examine the IR nature of galaxies.


\subsection{UV and IR luminosities.}\label{UV_FIR}

      \begin{figure}
   \centering
   \includegraphics[width=0.47\textwidth]{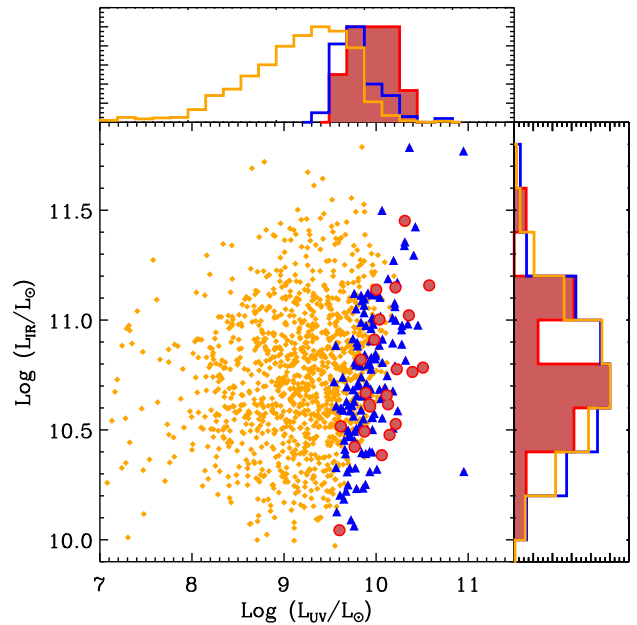}
   \caption{Total IR luminosities against UV luminosity for LAEs (red points), non-LAEs (blue triangles) and UV-faint galaxies (yellow squares). Histograms for each kind of galaxy are also shown with the same color code.
              }
         \label{histo_lumi}
   \end{figure}

$L_{\rm UV}$, expressed as $\nu L_{\nu}$, is obtained for each galaxy from its observed magnitude in the NUV band and using the assumed cosmology. We choose NUV band because at $z \sim 0.3$, it covers the continuum near the Ly$\alpha$ line and  is not contaminated by the Ly$\alpha$ emission. On the IR side, $L_{\rm IR}$ is defined as the integrated IR luminosity between 8 and 1000 $\mu$m in the rest-frame and is obtained by using different calibrations depending on whether an object is PACS-detected or PACS-undetected but MIPS--24$\mu$m-detected. In the first case, $L_{\rm IR}$ is calculated using calibrations between PACS bands and total IR luminosities (Eqns \ref{cal1} and \ref{cal2}):

\begin{equation}\label{cal1}
\log{L_{\rm IR}} = 0.99 \log{L_{100\,\mu{\rm m}}} + (0.44 \pm 0.25)
\end{equation}

\begin{equation}\label{cal2}
\log{L_{\rm IR}} = 0.96 \log{L_{160\,\mu{\rm m}}} + (0.77 \pm 0.21)
\end{equation}

\noindent where all the luminosities are in solar units and $L_{100\,\mu\textrm{m}}$ and $L_{160\,\mu\textrm{m}}$ are defined as $\nu L_{\nu}$. The calibration employed for each galaxy is that associated with the reddest PACS band where it is detected. This would represent the nearest measurement to the dust emission peak, ensuring a better determination of $L_{\rm IR}$. To derive these calibrations, we focus on the COSMOS field and select all the PACS-detected objects both at 100 $\mu$m and 160 $\mu$m, which are spectroscopically catalogued to be at $z \lesssim 0.5$ in the zCOSMOS survey \citep{Lilly2007}. This condition enables us to carry out accurate FIR SED fits with \cite{Chary2001} (hereafter CE01) templates since at that $z \lesssim 0.5$ the dust emission peak is well sampled with those FIR bands. The fits are performed with the Zurich Extragalactic Bayesian Redshift Analyzer \cite[ZEBRA,][]{Feldmann2006} in the maximum likelihood mode and the $L_{\rm IR}$ for each galaxy is obtained by integrating its best fitted template between 8 and 1000 $\mu$m in the rest-frame. The calibrations were built by comparing the $L_{\rm IR}$ of each source with that in the PACS bands and the errors are twice the standard deviation in the fittings. In this process, we ruled out AGN via X-ray emission diagnosis.

For the PACS-undetected but MIPS--24$\mu$m-detected objects, we convert MIPS--24$\mu$m fluxes into $L_{\rm IR}$ by fitting the fluxes to CE01 templates. These templates are built in such a way that for a given flux and redshift a unique solution for $L_{\rm IR}$ exists. The determination of $L_{\rm IR}$ from single FIR band extrapolations has also been employed in other studies \citep{Elbaz2010,Nordon2010,Elbaz2011}. \cite{Elbaz2010} analyze the applicability of this procedure, finding that MIPS--24$\mu$m extrapolations to $L_{\rm IR}$ are valid for galaxies up to $z \sim 1.5$ and which fall below the ULIRG limit. Both conditions are met in our case, both in LAEs and in the galaxies belonging to the control sample.

In Figure \ref{histo_lumi} we present the $L_{\rm UV}$ and $L_{\rm IR}$ for LAEs at $z \sim 0.3$ and those for the control sample, both non-LAEs and UV-faint galaxies. On the UV side, it can be seen that the histogram for LAEs includes larger values than that for non-LAEs, indicating that LAEs are UV-brighter than non-LAEs. Median $\log{(L_{\rm UV})}$ for LAEs and non-LAEs are 10.1 and 9.9, respectively. On the IR side, $L_{\rm IR}$ values for LAEs and non-LAEs are in the same range, mainly spanning from $\log{(L_{\rm IR}/L_{\odot})} = 10.4$ to 11.2. Even for UV-faint galaxies, the histogram of $L_{\rm IR}$ is similar to that for LAEs and non-LAEs, despite their difference in the UV selection.


\subsubsection{IR emission and ULIRG fraction evolution}\label{IRnature}

According to their $L_{\rm IR}$, galaxies can be classified into three types: normal SF galaxies: $L_{\rm IR} < 
10^{11}\,L_{\odot}$, luminous infrared galaxies (LIRGs): $10^{11}< L_{\rm IR}$ [$L_{\odot}$] $<$ 10$^{12}$ and ultra-luminous infrared galaxies (ULIRGs): $L_{\rm IR}> 10^{12}\,L_{\odot}$. As  can be seen in Figure \ref{histo_lumi}, most of our IR-detected LAEs are in the normal SF regime. Due to the correlation between MIPS--24$\mu$m luminosity and $L_{\rm IR}$, IR-undetected LAEs with FIR coverage have $L_{\rm IR}$ less than the IR limiting luminosity ($\sim$10$^{10.4}\,L_{\odot}$), being normal SF galaxies too. In this way, considering the 30 LAEs with IR coverage (both IR-detected and IR-undetected), we find that at $z\sim 0.3$ more than 80\% of LAEs are SF galaxies. Only six LAEs have $10^{11} < L_{IR}[L_{\odot}] < 10^{11.5}$, and none of them have $L_{\rm IR}> 10^{11.5}$ nor fall in the ULIRG class. Note that, considering both IR-detected and undetected LAEs, we work with a nearly complete sample of LAEs up to $m_{\rm NUV}\sim 21.5$ mag \citep{Cowie2010}, and therefore, the non-existence of ULIRGs in our sample of LAEs is an unbiased result under that limit. If there were LAEs with an ULIRG nature at $z\sim 0.3$, they would have to be fainter in the UV than those analyzed in the present work. No ULIRG is found in the control sample (non-LAEs and UV-faint galaxies) either, most galaxies being in the normal SF regime as well. Indeed, most recent studies of the FIR properties of galaxies with \emph{Herschel} do not find many ULIRGs at $z\sim 0.3$, but they begin appearing from $z\sim 1.0$ \citep{Elbaz2011,Elbaz2010,Buat2010,Lutz2011}.

At $z \gtrsim 2$, \cite{Chapman2005} found Ly$\alpha$ emission in the optical spectra of a sample of 850 $\mu$m detected SF sub-mm galaxies with a ULIRG nature. \cite{Nilsson2009_letter,Nilsson2011} also suggest that some LAEs at $z \sim 2$ have a ULIRG nature. In this way, the discovery of Ly$\alpha$ emission in ULIRGs at $z \gtrsim 2$ and the fact that most of our LAEs at have $L_{\rm IR} \lesssim 11.2 L_\odot$ suggests that the IR nature of the galaxies found via their Ly$\alpha$ emission is changing from $z \sim 2$ to $z \sim 0.3$: there is a population of red and dusty LAEs with a ULIRG nature at $z \gtrsim 2$, which is not seen at $z \sim 0.3$. This trend is similar to that found in previous studies for general galaxies detected in MIPS--24$\mu$m and to the trend of cosmic star formation \citep{Hoog1998, Hopkins2004, Hopkins2006,Perez2005,Lefloch2005}. \cite{Cowie2011} find a dramatic evolution in the maximum of the Ly$\alpha$ luminosity function between $z \sim 0.3$ and $z \gtrsim 1.0$. This result, together with the evolution in the IR emission of LAEs  suggested above, indicates that either the properties of galaxies selected via their Ly$\alpha$ emission are evolving over cosmic time or the Ly$\alpha$ selection technique is not tracing the same kind of objects at different redshifts.


\subsection{Dust attenuation}\label{dust_content}

   \begin{figure*}
   \centering
   \includegraphics[width=0.8\textwidth]{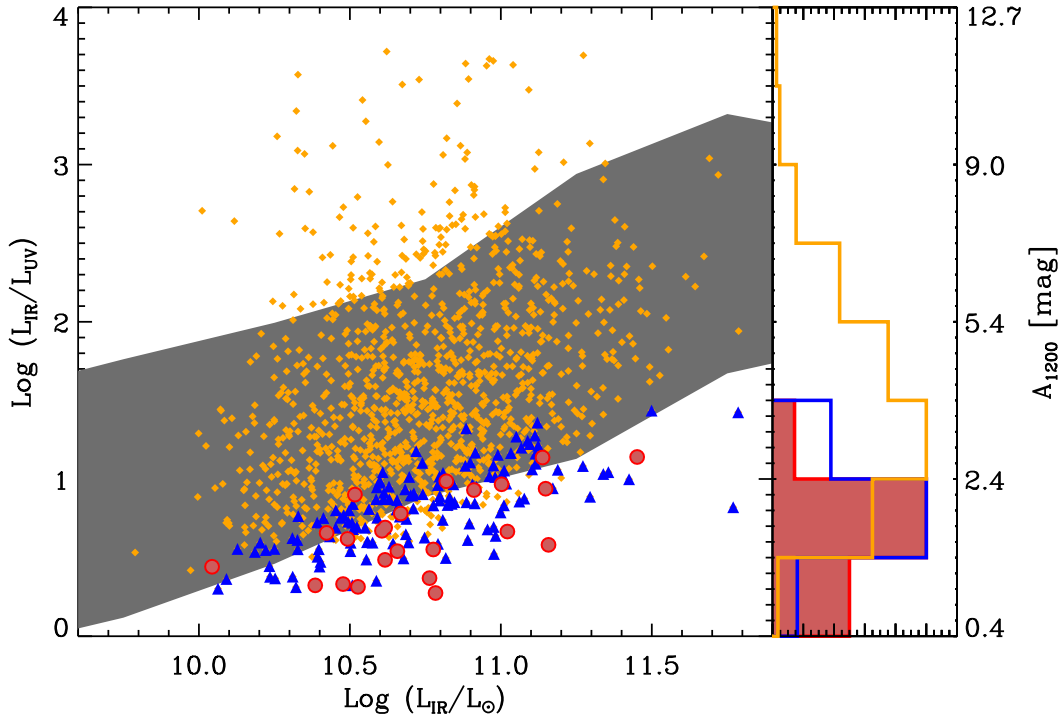}  
  \caption{Dust attenuation against $L_{\rm IR}$ for LAEs (red dots), non-LAEs (blue triangles) and UV-faint galaxies (yellow squares). We also represent in gray, as a comparison, the average zone ($\pm$$\sigma$) where the local galaxies of \cite{Gil2007} are located. The histogram of dust attenuation for each kind of galaxy is included in the right side of the plot, with the same color code. Histograms have been normalized to the maximum in order to clarify the representation.
              }
         \label{histo_dust}
   \end{figure*}

The ratio between $L_{\rm IR}$ and $L_{\rm UV}$ is a good tracer of dust attenuation in galaxies. Here, we adopt the calibration between the IR/UV ratio and dust attenuation found in \cite{Buat2005} (Equation \ref{A_NUV}) to obtain dust attenuation for our IR-detected LAEs at $z \sim 0.3$ and for the galaxies in the control sample, both non-LAEs and UV-faint galaxies:

\begin{equation}\label{A_NUV}
A_{\rm NUV} = -0.0495x^3 + 0.4718x^2 + 0.8998x + 0.169
\end{equation}

\noindent where $A_{\rm NUV}$ is the dust attenuation in the NUV band and $x=\log{\left(L_{\rm IR}/L_{\rm UV}\right)}$. The conversion from $A_{\rm NUV}$ to the dust attenuation at 1200\AA, $A_{1200\,\AA}$, is made by using the \cite{Calzetti2000} reddening law. As in \cite{Buat2007b}, we do not apply $K$-correction to the UV luminosities given that the $L_{\nu}$ spectrum of galaxies is quite flat in the UV range. In this way, we assume that the NUV fluxes observed at $z\sim 0.3$ is the same as those observed at $z\sim 0$ in the NUV band.

In Figure \ref{histo_dust} we represent the dust attenuation of our LAEs versus their $L_{\rm IR}$ along with the dust attenuation distribution, which has a median value of $A_{1200\AA}\sim 1.5$ mag. We also plot in that figure the dust attenuation for non-LAEs and for the UV-faint galaxies. It can be seen that, for each value of $L_{\rm IR}$, LAEs are slightly less dusty than non-LAEs. In fact, the median value of the dust attenuation distribution for non-LAEs is $A_{1200\AA}\sim 2.0$ mag. Furthermore, for both LAEs and non-LAEs there is a trend towards brighter galaxies in the IR being dustier.

\cite{Cowie2011} find, comparing the UV spectral index to the H$\alpha$/H$\beta$ flux ratio, that the dust attenuation in most LAEs at $z\sim 0.3$ can be also described by \cite{Cardelli1989} or \cite{Fitzpatrick1999} laws, i.e., the stellar extinction is better represented as a uniform screen rather than a patchy distribution. The inclusion of those laws here would change the final numbers in the sense that the dust attenuation would be lower than that obtained with the \cite{Calzetti2000} law, although the relation between the dust attenuation between different populations would remain the same.

LAEs with a fainter UV continuum than the limiting magnitude are not included in the sample and they could be dustier than those presented in this study. Therefore, there could be some LAEs at $z\sim 0.3$ which are dustier and have larger rest-frame Ly$\alpha$ EW equivalent widths than those presented here. The finding of a Ly$\alpha$ emission in the UV spectra of a dusty object at $z\sim 0.3$ could indicate the presence of a clumpy ISM \citep{Neufeld1991}. The possible existence of this and other kinds of geometries has been reported to be able to recover the observed Ly$\alpha$/H$\alpha$ and H$\alpha$/H$\beta$ ratios and the observed SEDs of LAEs at different redshifts \citep{Finkelstein2011_espectros,Scarlata2009,Guaita2011,Finkelstein2008}.

\subsubsection{Dust attenuation and UV continuum slope}

      \begin{figure}
   \centering
   \includegraphics[width=0.47\textwidth]{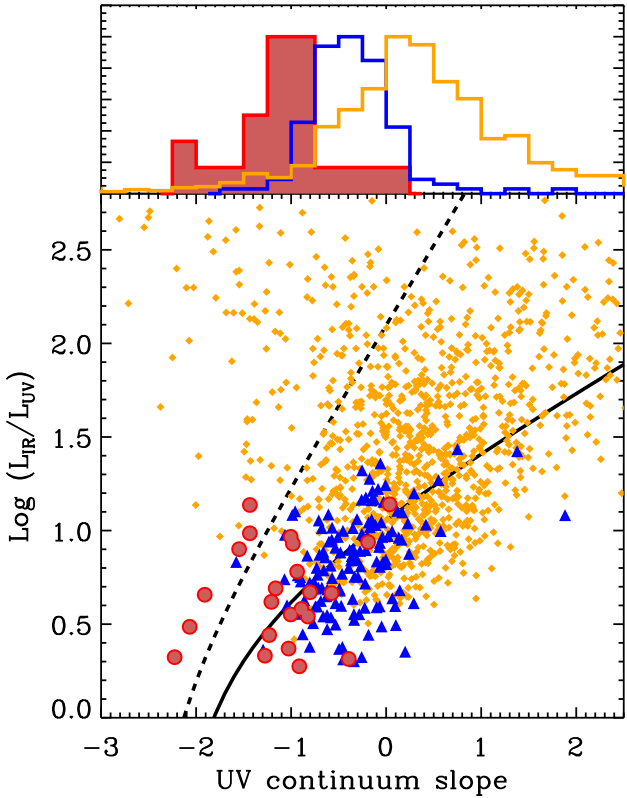}
   \caption{Dust attenuation against the UV continuum slope ($\beta$) as obtained with GALEX photometry for LAEs (red dots), non-LAEs (blue triangles) and UV-faint galaxies (yellow squares). Solid and dashed lines show the relation for local normal SF galaxies and local SBs, respectively. The distribution of the UV continuum slope for each kind of galaxy is shown in the upper plot, with the same color code. Histograms have been normalized to the maximum in order to clarify the representation.
              }
         \label{dust_beta}
   \end{figure}

Dust attenuates the rest-frame UV light of a galaxy in a wavelength-dependent way, this attenuation being larger for shorter UV wavelengths. Therefore, dust produces an increase in the rest-frame UV continuum slope, $\beta$, from negative to less negative or even positive values. On the other hand, part of the absorbed light in the UV is in turn re-emitted in the FIR regime. In this way, dust attenuation and $\beta$ are expected to be related. Indeed, many studies employ different relations between dust attenuation and $\beta$ to obtain the dust attenuation in galaxies that are not detected in the FIR.

In this section we compare dust attenuation and $\beta$ for the IR-detected LAEs, non-LAEs, and UV-faint galaxies. Shown in Figure \ref{dust_beta} is the IRX-$\beta$ diagram for the three kinds of galaxies. $\beta$ values are obtained by employing the \cite{Kong2004} recipe, as in \cite{Buat2010}. In this way, we assume that the rest-frame UV continuum can be described by a power law in the form $f_{\lambda}\sim \lambda^\beta$ and obtain $\beta$ from GALEX photometry in the FUV and NUV channels. It can be clearly seen that LAEs have UV continuum slopes much bluer than non-LAEs, which is compatible with their being among the least dusty galaxies at their redshift (see Section \ref{dust_content}). The finding of larger (redder) values of $\beta$ for the UV-faint galaxies than for LAEs and non-LAEs is due to their faintness in the UV.

Shown in Figure \ref{dust_beta} are also the curves of \cite{Kong2004} and \cite{Boissier2007} corresponding to normal SF and starburst (SB) galaxies, respectively. It can be seen that although many LAEs are distributed around the curve corresponding to the normal SF galaxies, some of them occupy a different location in the IRX-$\beta$ diagram, quite near to the SB relation. Some studies use a relation between dust attenuation and $\beta$ to obtain the dust attenuation from the values of the UV continuum slope for objects which are not detected in the FIR. The reason for this procedure is that, whereas the UV continuum slope is relatively easy to obtain from broad-band photometry over a wide range of redshifts, the detection rate in the FIR of certain kinds of galaxies is very low, mainly at the highest redshifts. This is typically the case for LAEs \citep{Oteo2010_z2}. But it should be noted that the reliability of this technique is based on that the relation assumed between dust attenuation and $\beta$ applies for the population of galaxies under study. What we find here is that this is not the case for LAEs at $z \sim 0.3$, which are distributed between the SB and the SF relations so that a unique relation cannot be applied to the whole population to recover the dust attenuation from $\beta$. Furthermore, it has been reported that the location of galaxies in the IRX-$\beta$ diagram also depends on the $L_{\rm IR}$ and age. For example, at $z \sim 2$, galaxies with ages below 100 Myr tend to be less reddened at a given UV continuum slope than the values predicted by the SB relation \citep{Reddy2006,Reddy2010,Reddy2012}. This tendency might have a significant influence on LAEs, since most of them have been reported to be young galaxies in a wide range of redshifts.

On the other hand, non-LAEs are mostly distributed around the normal SF curve, although in a zone associated with higher $\beta$ values than those for LAEs. There is a very low percentage of non-LAEs near the SB curve. From another point of view, for each value of dust attenuation, LAEs tend to have bluer UV continuum slopes than non-LAEs. This could indicate a different star formation mode between LAEs and non-LAEs.

\subsubsection{Dust attenuation and redshift}

   \begin{figure}[!t]
   \centering
   \includegraphics[width=0.47\textwidth]{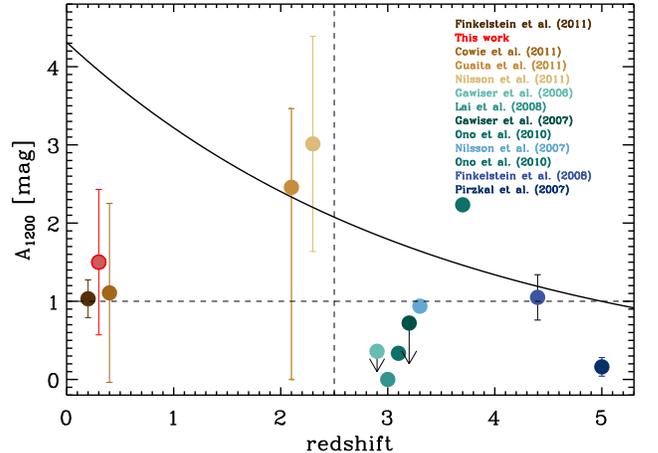}
   \caption{Dust attenuation of LAEs at different redshifts reported in this (red dot) and previous studies (brown, green, and blue dots). Dots with error bars indicate that the results have been obtained examining the objects individually and, therefore, a distribution of values can be obtained. In those cases, the error bars represent the standard deviation of the distributions. Dots without error bars indicate that the results have been obtained by using a stacking analysis so that one value represents the average behavior of the sample. Some points have been slightly shifted in redshift in order to clarify the representation. Vertical and horizontal dashed lines are represented to clarify the suggested evolution in dust attenuation between $z \lesssim 2.5$ and $z \gtrsim 2.5$, as explained in the text. The solid curve represents the redshift evolution of dust attenuation reported in \cite{Hayes2011}.
              }
         \label{dust_z}
   \end{figure}

Two open questions in the study of LAEs are their dust attenuation and its possible evolution with redshift. In the previous sections we provided additional clues to answer the former question for LAEs at $z\sim 0.3$ by using UV and mid-IR/FIR measurements. Now, in an attempt to answer the latter question, we compare our values with those at different redshifts reported in previous studies \citep{Finkelstein2011_espectros,Cowie2011,Guaita2011,Nilsson2011,Gawiser2006, Lai2008,Gawiser2006, Ono2010,Nilsson2007,Finkelstein2008, Finkelstein2009d, Pirzkal2007}. Figure \ref{dust_z} shows the representation of the results reported in those works, along with the curve of the redshift evolution of a general population of galaxies found in \cite{Hayes2011}. We also include the line associated with a dust attenuation of $A_{1200\AA}$ = 1.0 mag at all redshifts in order to clarify the differences in dust attenuation between LAEs at different redshifts. It can be seen that most LAEs at $z\gtrsim 2.5$ have dust attenuation below 1.0 mag in 1200 \AA. Actually, most results indicate that the SEDs of LAEs at those redshifts are compatible with dust attenuation below 0.5 mag at 1200 \AA \citep{Lai2008,Gawiser2006,Gawiser2007}.

At $z\sim 0.3$, we find a dust attenuation distribution centered in $A_{1200\,\AA}$ = 1.5 mag. This is compatible with the results reported by \cite{Finkelstein2011_espectros} and \cite{Cowie2011} even though the results were obtained with rest-frame optical spectroscopy and SED fitting and consequently different methodologies. 

At $z \sim 2.0$ \cite{Guaita2011} and \cite{Nilsson2011} report high dust attenuation values, higher than those reported at $z \sim 0.3$ and $z \gtrsim 2.5$. Although these median values are high, the dust attenuation distributions found in both studies are quite wide, which makes it difficult to establish a clear tendency between $z \sim 2.0$ and other redshifts. However, the high median values themselves are an indication that there should exist a population of dusty LAEs at $z \sim 2.0$ that are not seen at other redshifts. Furthermore, the redshift distribution of the red and dusty objects selected via their sub-mm emission is centred around $z \sim 2.3$ \citep{Chapman2005}. Some of those galaxies are SF ULIRGs exhibiting  Ly$\alpha$ emission in the spectra, which reinforces the idea that, at those redshifts, the Ly$\alpha$ selection technique can also segregate dusty galaxies, in contrast with the classical idea of LAEs as galaxies with low dust attenuation. \cite{Blanc2011} find the emergence of a small population of red galaxies at $z < 3$ (in terms of their UV continuum slope) in their study of integral-field espectroscopically selected LAEs at $1.9 < z < 3.8$. Moreover, in \cite{Oteo2010_z2} we report the detection in PACS--160$\mu$m of a sample of LAEs at $2.0 \lesssim z \lesssim 3.5$ and derive their dust attenuation by employing the IR/UV ratio \citep{Buat2005}. As result, we obtain values of $A_{1200\,\AA} \gtrsim 4.5$ mag, higher than the median values reported in \cite{Guaita2011} and \cite{Nilsson2011} but compatible with the width of the distribution in \cite{Nilsson2011}. Note that the values of dust attenuation reported in \cite{Oteo2010_z2} represent the upper limit on LAEs at that redshift, since we only work with FIR-detected sources.

At $z \sim 3.1$, most LAEs are found to be nearly dust-free objects \citep{Gawiser2006,Gawiser2007,Lai2008}, representing a great difference from $z\sim 3.1$ to $z \sim 2.1$--2.3. However, it should be considered that at $z\gtrsim 2.5$, the dust attenuation in most studies has been obtained by using a stacking analysis of the sample. \cite{Nilsson2011} compared the dust attenuation obtained for LAEs at $z \sim 2.3$ when performing a stacking analysis with those found when considering individual objects. They found that dust attenuation is considerably reduced when stacking the total sample and a subsample of old LAEs, indicating that dust attenuation is very sensitive to stacking.

Despite the uncertainties of stacking, the evolution of dust attenuation between $z\gtrsim 3$ and $z\sim 2$ could  be related to the evolution in other properties of LAEs. \cite{Ciardullo2012} report that there is a significant evolution in the LAE luminosity function between $z \sim 3.1$ and $z \sim 2.1$. \cite{Bond2011_evolution} claim for an evolution in the size of LAEs from $z\sim 3.1$ to $z\sim 2.1$, raising the median half-light radius from 1.0 kpc at $z\sim 3.1$ to 1.4 kpc at $z\sim 2.1$. \cite{Nilsson2009} also found evolution in the physical properties of LAEs between $z\sim 3$ and $\sim$2, in the sense that at $z\sim 2$ LAEs have redder SEDs, which indicates that they are more evolved (dustier, older and more massive) than those at $z\sim 3$. 

At $z\gtrsim 4$, the samples studied contain very few objects and the results are not as statistically significant as at lower redshifts; therefore, although most studies have reported that the SED of LAEs is compatible with some levels of dust, more work is needed to confirm that behavior.

Therefore, the general trend is that LAEs at $z\gtrsim 2.5$ tend to be slightly less dusty than their low-redshift analogues. Furthermore, it can be seen that LAEs at $z\gtrsim 2.0$ tend to follow the dust evolution of galaxies with redshift, while at $z\sim 0.3$, they deviate from that behavior.

Apart from indicating a possible evolution of dust attenuation with redshift for LAEs, Figure \ref{dust_z} also shows the lack of knowledge that we still have about the physical properties of those galaxies. While the best method of determining dust attenuation is the combination of direct UV and FIR measurement used in this study, different procedures have been used to derive that parameter. However, it has not been checked yet whether they provide consistent results at a given redshift. Furthermore, at the highest redshifts, either the number of objects studied is low, or stacking analysis had to be done due to the non-detection of LAEs in many photometric bands.

   
\subsection{Star Formation Rates}\label{SFR}

   \begin{figure}
   \centering
   \includegraphics[width=0.47\textwidth]{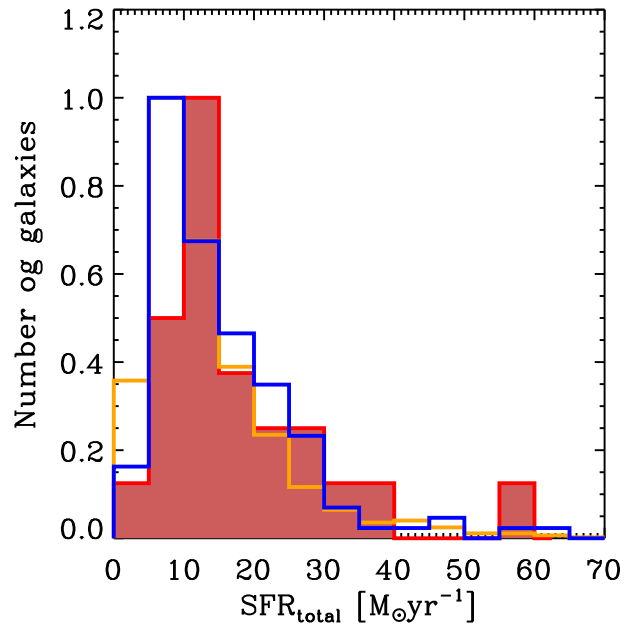} \\
   \caption{Distribution of the SFR$_{\rm total}$ = SFR$_{\rm UV,uncorrected}$ + SFR$_{\rm IR}$ for our LAEs (red-shaded), non-LAEs (blue), and other UV-faint galaxies (yellow). Histograms have been normalized to the maximum in order to clarify the representation.
              }
         \label{histo_SFR}
   \end{figure} 

The combination of UV and IR data also provides the most accurate determination of SFR in galaxies. Adopting the \cite{Kennicutt1998} calibrations, the SFR associated with the observed UV and IR luminosities are given by the expressions:

\begin{equation}\label{SFR_UV}
\textrm{SFR}_{\rm UV,uncorrected}[M_{\odot}\textrm{yr}^{-1}] = 1.4\cdot10^{-28}L_{\nu,\rm observed}
\end{equation}

\begin{equation}\label{SFR_IR}
\textrm{SFR}_{\rm IR}[M_{\odot}\textrm{yr}^{-1}] = 4.5\cdot10^{-44}L_{IR}
\end{equation}

\noindent where $L_{\nu,\rm observed}$ is in units of erg/s/Hz and $L_{\rm IR}$ is defined in the same way as in Section \ref{UV_FIR}. In order to obtain the SFR$_{\rm total}$ for our galaxies we assume that all the light absorbed in the UV is in turn re-radiated in the FIR. In this scenario, SFR$_{\rm total}$ can be calculated as the sum of a dust-uncorrected component, SFR$_{\rm UV,uncorrected}$ and the correction term shown in Eqn.\ \ref{SFR_IR}. Thus, it can be written:

\begin{equation}\label{SFR_total}
\textrm{SFR}_{\rm total} = \textrm{SFR}_{\rm UV, uncorrected} + \textrm{SFR}_{\rm IR}
\end{equation}

The distributions of the SFR$_{\rm total}$ for IR-detected LAEs and the galaxies in the control sample are shown in Figure \ref{histo_SFR}. It can be seen that IR-detected LAEs have SFR$_{\rm total}$, ranging mainly from 10 to 40 $M_{\odot}$ yr$^{-1}$, with a median of 18 $M_{\odot}$ yr$^{-1}$, and peaking around 13 $M_{\odot}$ yr$^{-1}$. The median value of SFR$_{\rm total}$ for non-LAEs is 15 $M_{\odot}$ yr$^{-1}$, with a distribution peaking around 8 $M_{\odot}$ yr$^{-1}$, which tends to be shifted to lower values of SFR$_{\rm total}$ than that for LAEs. The median value found in this study is larger than that reported in \cite{Cowie2011} by using extinction-corrected H$\alpha$ luminosities, SFR$_{\rm H\alpha-corrected}\sim 6 M_{\odot}$ yr$^{-1}$.

\begin{figure}
   \centering
   \includegraphics[width=0.47\textwidth]{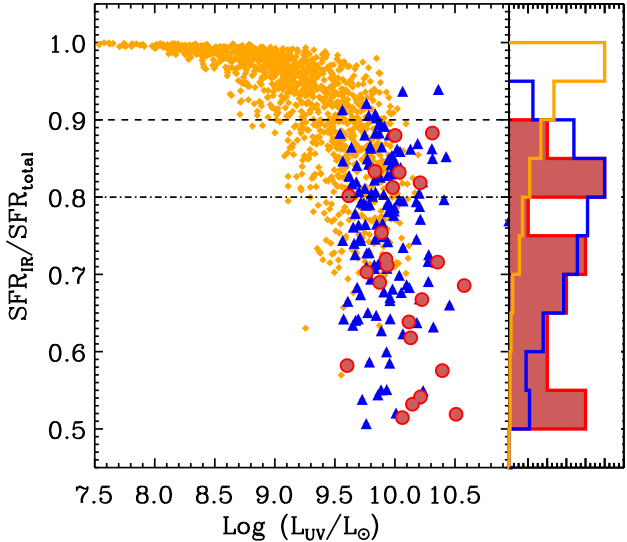} \\
   \caption{The ratio between SFR$_{\rm IR}$ and SFR$_{\rm total}$ against UV luminosity for our LAEs (red dots), non-LAEs (blue triangles), and other UV-faint galaxies (yellow squares). Dashed and dashed-dotted horizontal lines limit the zones where the SFR$_{\rm IR}$ represent 90\% and 80\% of SFR$_{\rm total}$, respectively.The distribution of SFR$_{\rm IR}$/SFR$_{\rm total}$ is also represented for the three kinds of galaxies, with the same color code. Histograms have been normalized to the maximum to clarify the representation.
              }
         \label{SFR_UV_FIR}
   \end{figure}

Figure \ref{SFR_UV_FIR} shows the ratio between SFR$_{\rm IR}$ and SFR$_{\rm total}$ against $L_{\rm UV}$ and $L_{\rm IR}$ for LAEs and the general population of galaxies at $z\sim 0.3$. Considering the galaxies in the control sample, it can be seen that the ratio has a strong dependence on NUV luminosity: the IR contribution is lower with increasing $L_{\rm UV}$. In Section \ref{counterparts} we found that UV-bright galaxies at $z\sim 0.3$ are more like those detected in the FIR than UV-faint ones. Now, we have found that, for IR-detected galaxies, UV-bright ones have a lower IR contribution to SFR$_{\rm total}$ (are more transparent) than UV-faint ones. There is a limit in the NUV up to which SFR$_{\rm IR}$ is a good estimator of SFR$_{\rm total}$. Considering ratios of 80\% and 90\%, SFR$_{\rm IR}$ is a good indicator of SFR$_{\rm total}$ for galaxies at $z\sim 0.3$ with  $L_{\rm NUV}\lesssim9\,L_{\odot}$ and $\lesssim 9.5L_{\odot}$, respectively. IR-detected LAEs at $z\sim 0.3$ are all above the mentioned thresholds; therefore, both the UV and IR light contribute significantly to SFR$_{\rm total}$. Despite this, the FIR contribution to SFR$_{\rm total}$ for most LAEs is greater than 60\%, indicating that rest-frame UV-based methods would underestimate SFR$_{\rm total}$ by a factor greater than two. This is an indication of the great importance of 
FIR measurements when calculating the SFR: even in galaxies with low dust attenuation, such as LAEs, FIR emission makes a significant contribution; therefore, FIR data must be used to obtain accurate results.

\subsection{Morphology}\label{morfology}

In the previous sections, we have analyzed the physical properties of LAEs with direct IR and UV measurements. We now  examine whether there is a difference between LAEs and non-LAEs regarding their morphology in the optical side of their SED. In order to carry out a precise morphological analysis, high spatial resolution images (i.e., HST/ACS) are required. In the case of non-LAEs, given that they are all located in the COSMOS field, there is ACS information for almost all of them, 127 (about 95\%) to be precise. However, the IR-detected LAEs are distributed in different fields, some of which do not have available HST images. This limits the morphological study to eight IR-detected LAEs.

\begin{figure}
   \centering
   \includegraphics[width=0.47\textwidth]{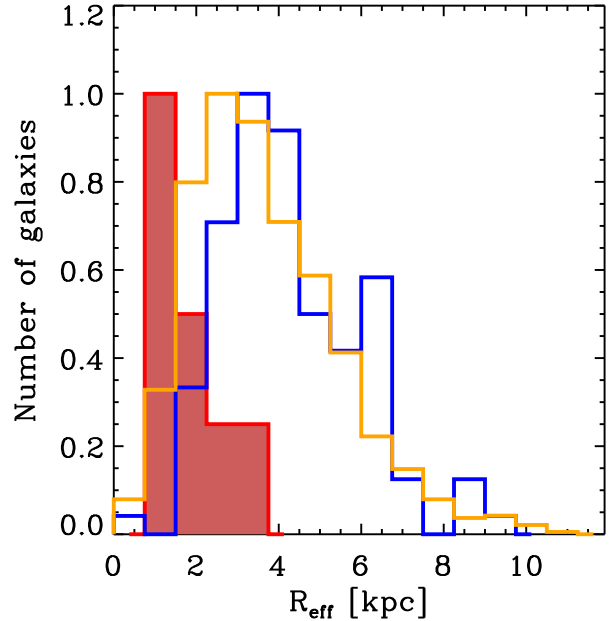} \\
   \caption{Distribution of the effective radius for our LAEs (red-shaded), non-LAEs (blue,) and other UV-faint galaxies (yellow). Histograms have been normalized to the maximum in order to clarify the representation.
              }
         \label{sizes}
   \end{figure}

We take $12''\times 12''$ $I$-band ACS cut-outs of our sources (LAEs and control sample) and follow both an analytical and a visual procedure. The analytical one is aimed at obtaining the physical sizes of the galaxies, whereas the visual one has the objective of classifying the galaxies within the different types of the Hubble sequence. Note that we use the images of the galaxies in the same passband and, since they are all in the same redshift range, we are analyzing the same zone of their SED, with no need for morphological $K$-correction. In order to obtain the sizes of our galaxies we fit Sersic profiles \citep{Sersic1968} to their light distributions by using GALFIT \citep{Peng2010}. Figure \ref{sizes} shows the light-half radii of LAEs and the galaxies in the control sample. It can clearly be seen that there is a remarkable difference: LAEs tend to be significantly smaller than non-LAEs. The median values of $R_{\rm eff}$ for LAEs, non-LAEs, and UV-faint galaxies are 1.5, 4.1, and 3.4 kpc, respectively. Note that non-LAEs are bigger than UV-faint galaxies owing to the difference in their UV brightness. The difference in size between LAEs and other UV-selected galaxies is also present at higher redshifts. \cite{Malhotra2011} report that LAEs tend to be smaller than LBGs at the same redshift from $z\sim 2$ up to $z\sim 4$. From $z\gtrsim 5$, LAEs and LBGs show similar sizes and have similar properties.

At $z\gtrsim 2$, narrow-band selected LAEs have been reported to be compact \citep{Bond2009,Bond2011_evolution, Malhotra2011}. \cite{Bond2009}, and \cite{Bond2011_evolution} report median values for $R_{\rm eff}$ of 1.0 and 1.4 kpc for their LAEs at $z\sim 3.1$ and $\sim$2.1, respectively. The median value for LAEs that we obtain in this study is not significantly different from the values reported at higher redshifts; hence, a clear evolution in the median value of the physical size of LAEs is not seen from $z\sim 2$ down to $z\sim 0.3$.

\begin{figure*}
   \centering
   \includegraphics[width=0.22\textwidth]{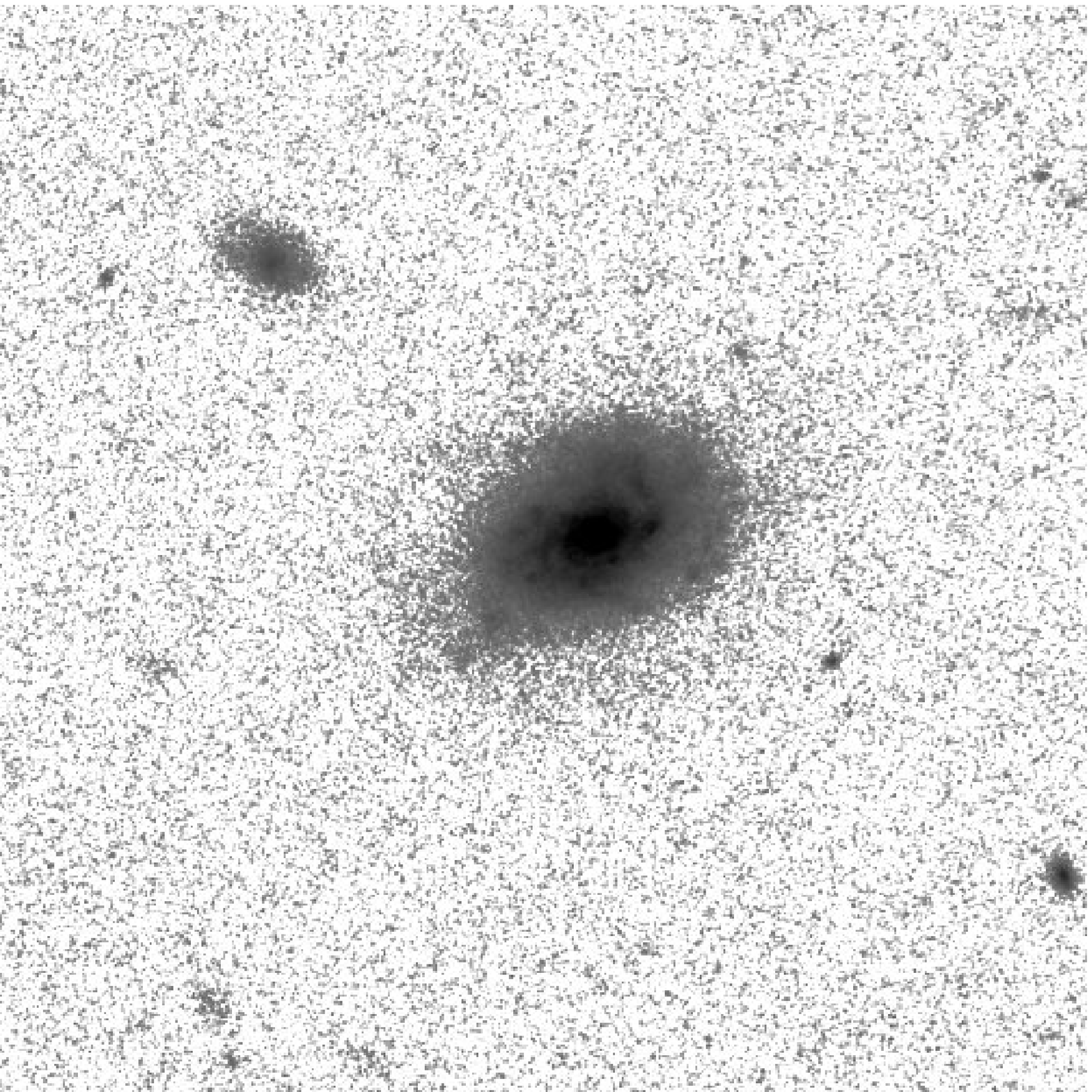} 
   \includegraphics[width=0.22\textwidth]{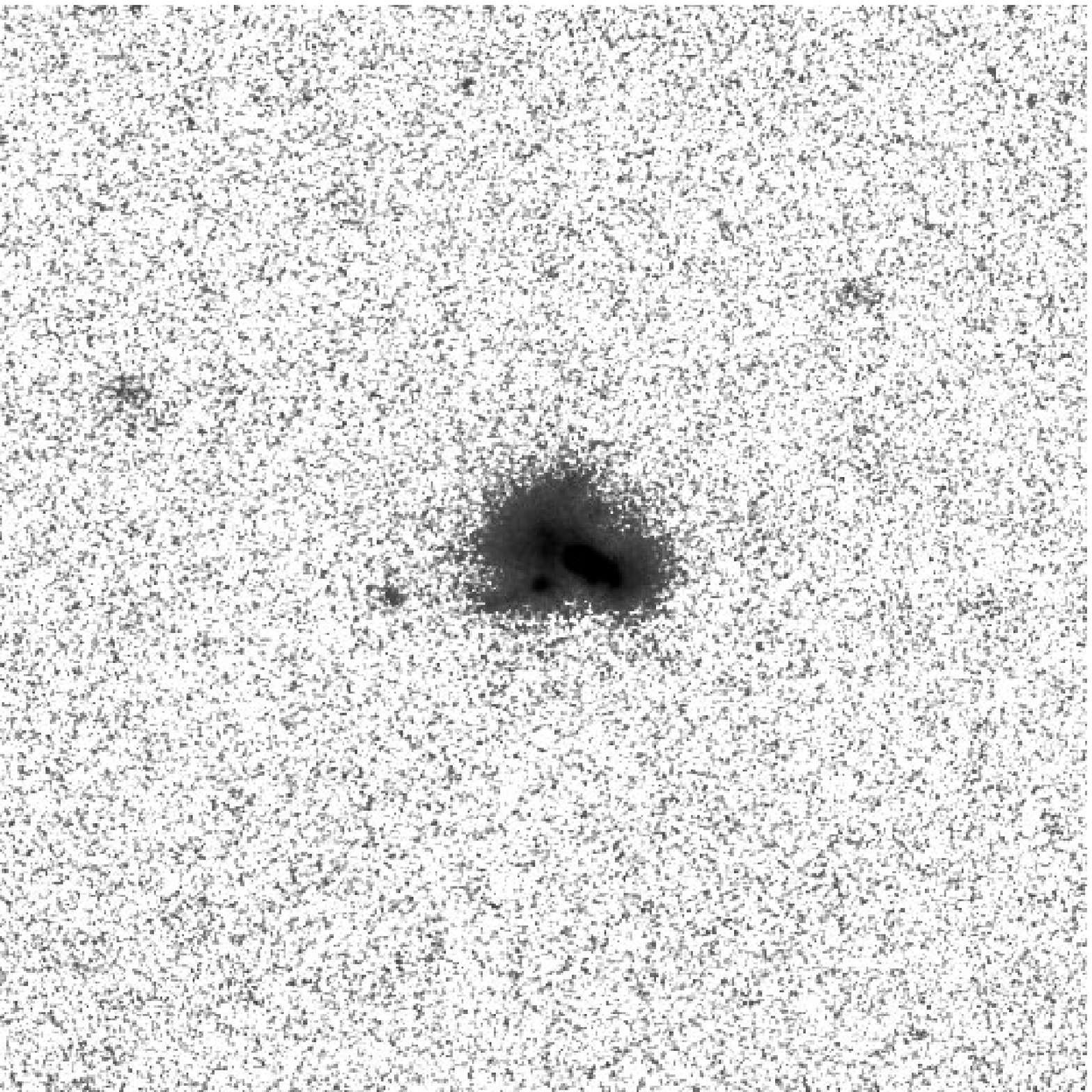} 
   \includegraphics[width=0.22\textwidth]{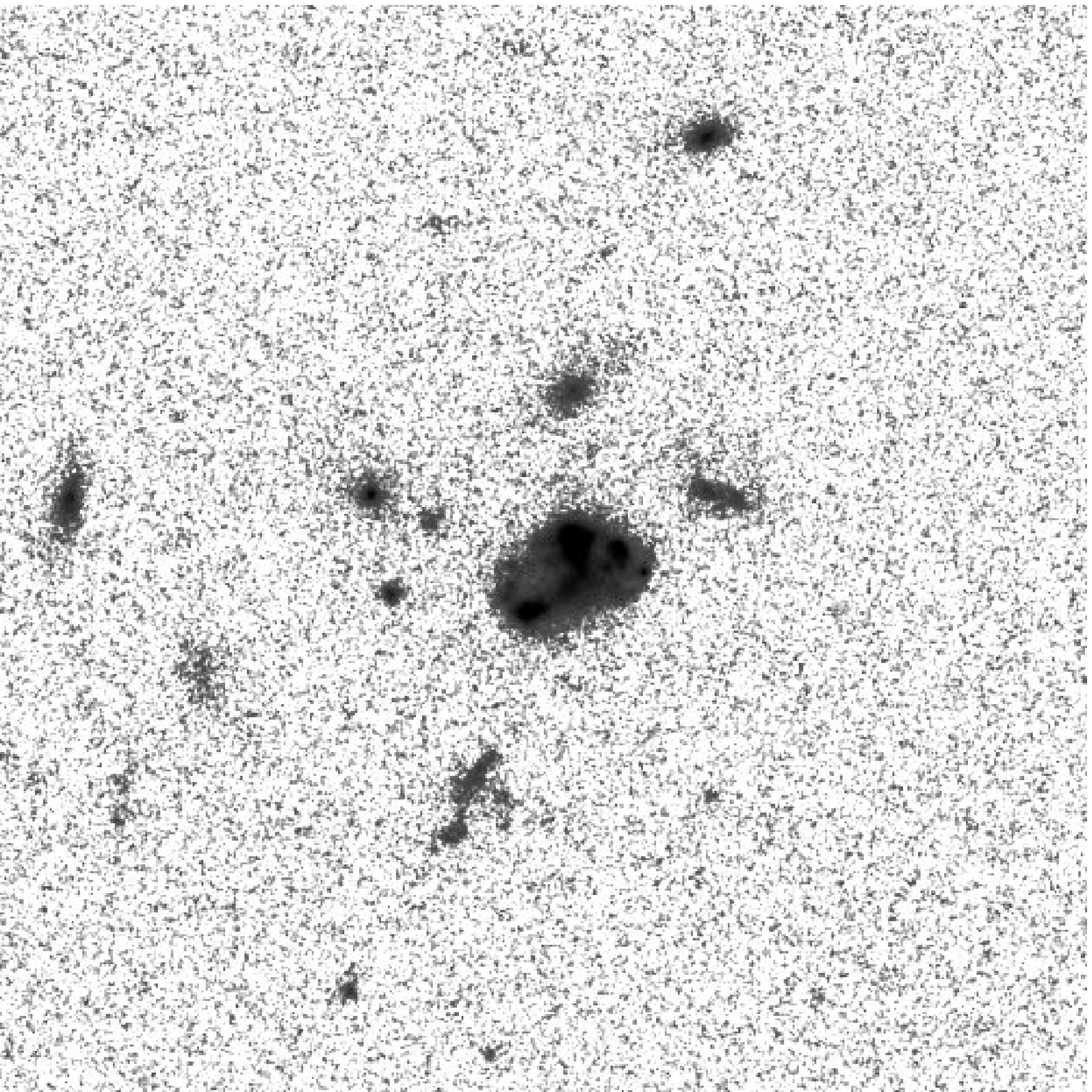} 
   \includegraphics[width=0.22\textwidth]{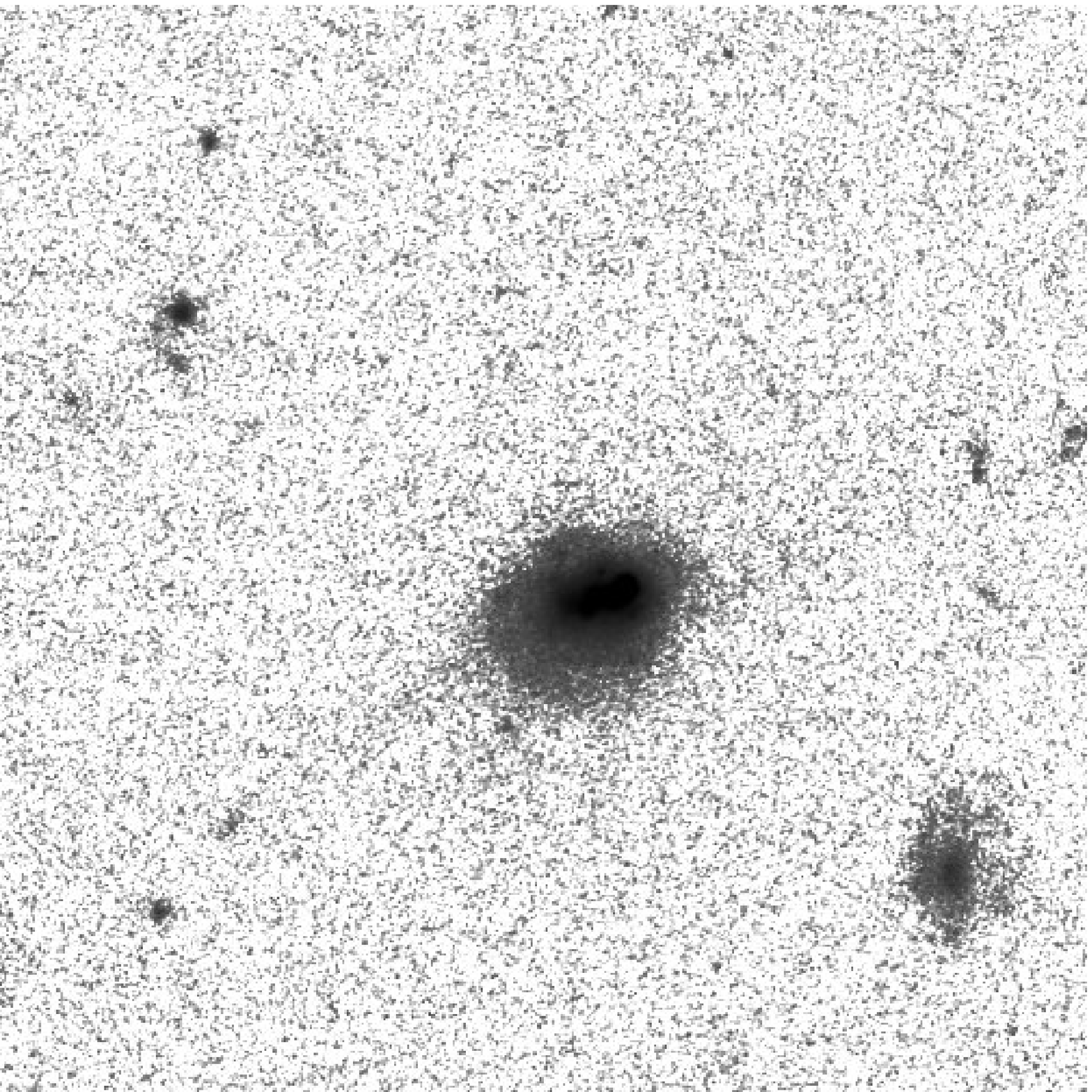} \\

   \includegraphics[width=0.22\textwidth]{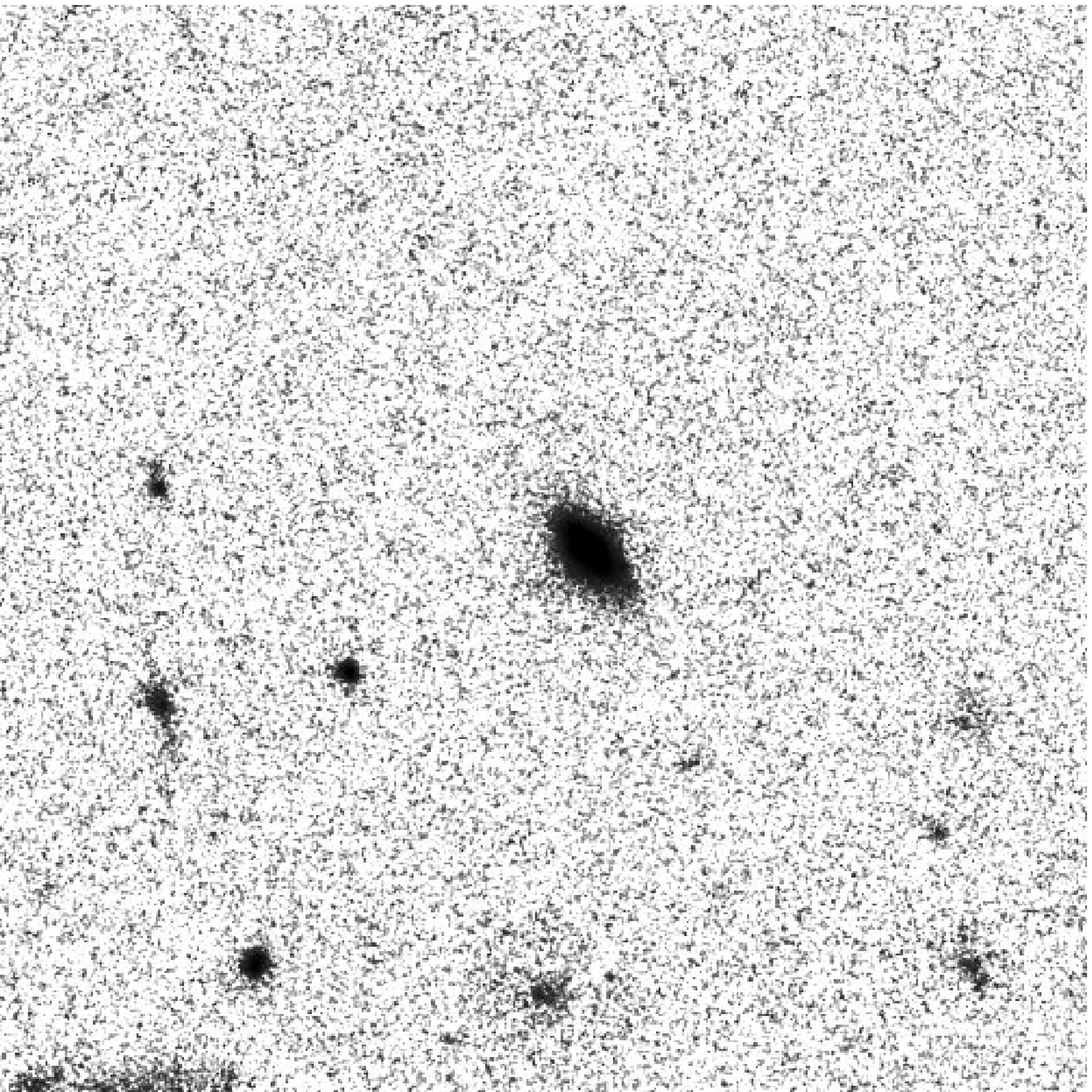}
   \includegraphics[width=0.22\textwidth]{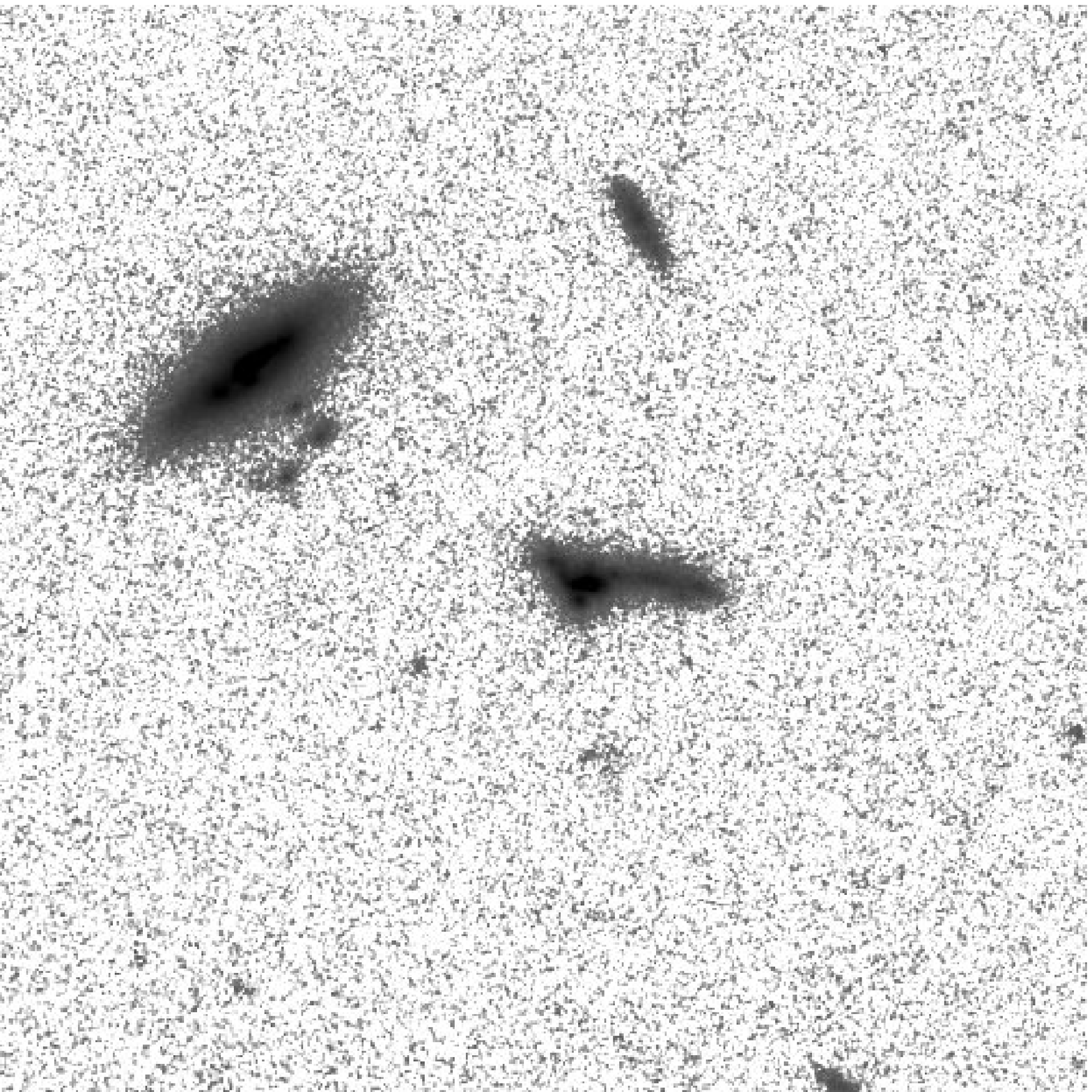} 
   \includegraphics[width=0.22\textwidth]{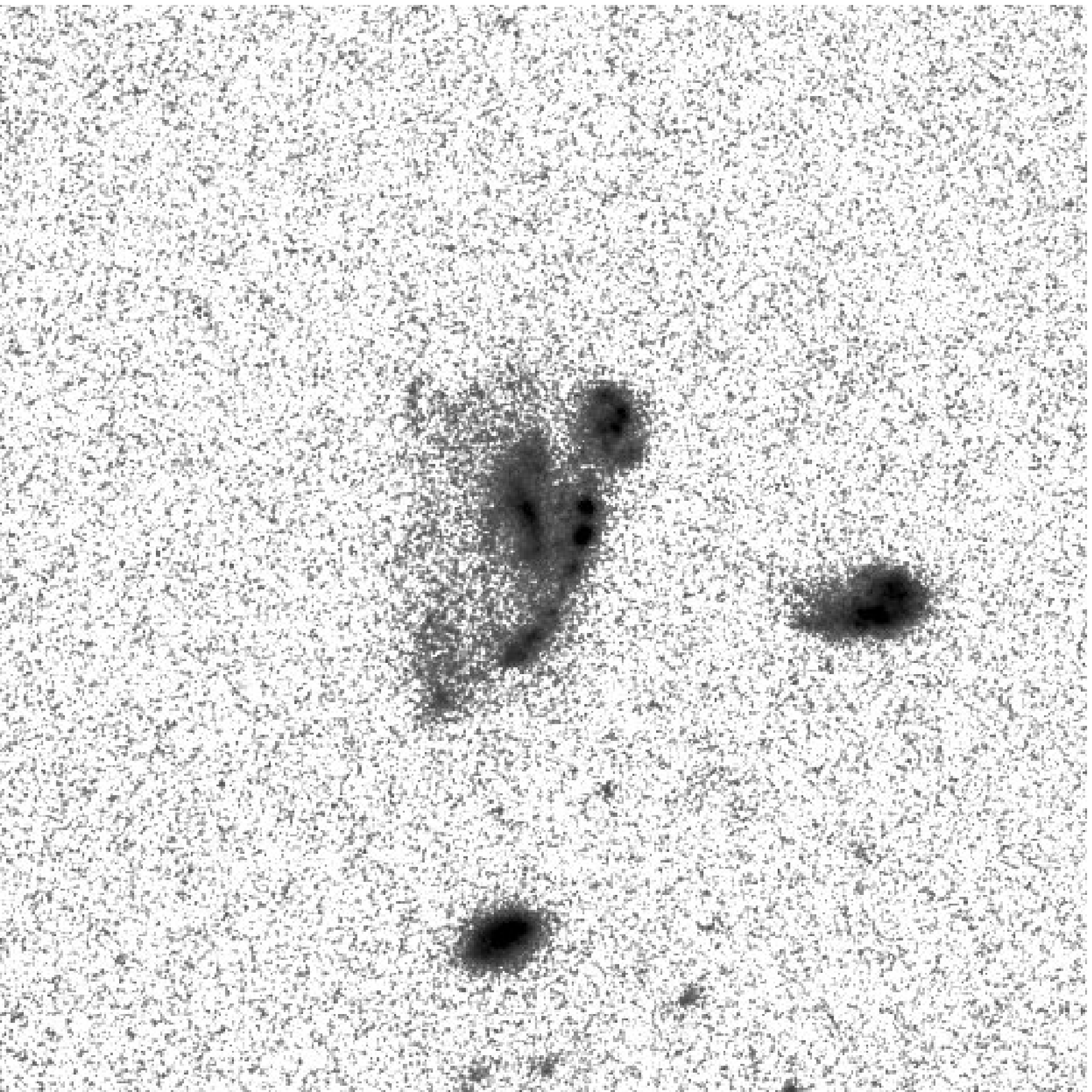}
   \includegraphics[width=0.22\textwidth]{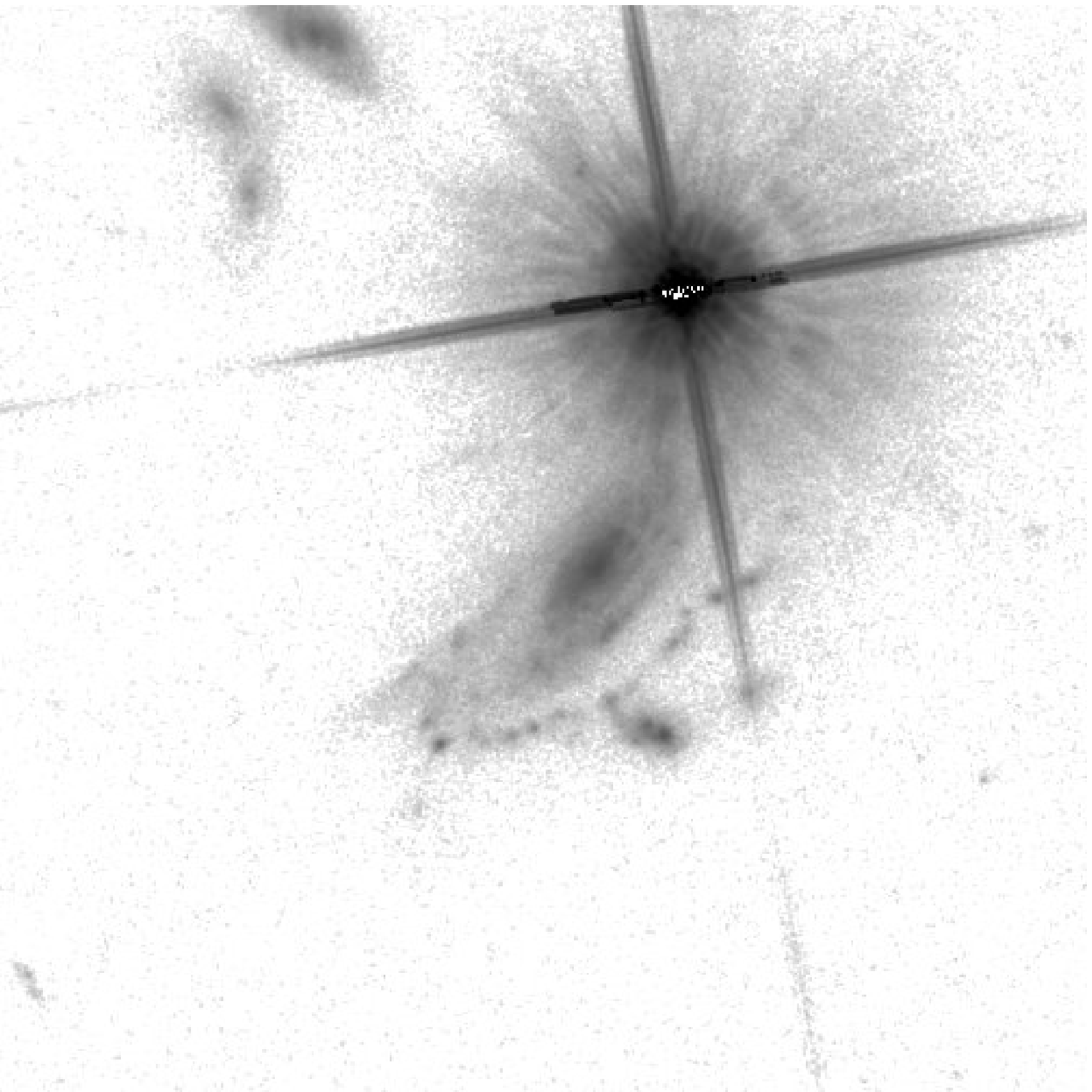} 
   \caption{ACS I-band 12''x12'' cut-outs of the 8 IR-detected LAEs with available ACS observations. LAEs are the galaxies in the center of each image. At z$\sim$0.3, the scale is 4.45 kpc/'', which means that the diameter of each cut-out represents 53.4 kpc.
              }
         \label{shapes_LAEs}
   \end{figure*}

\begin{figure*}
   \centering
\includegraphics[width=0.9\textwidth]{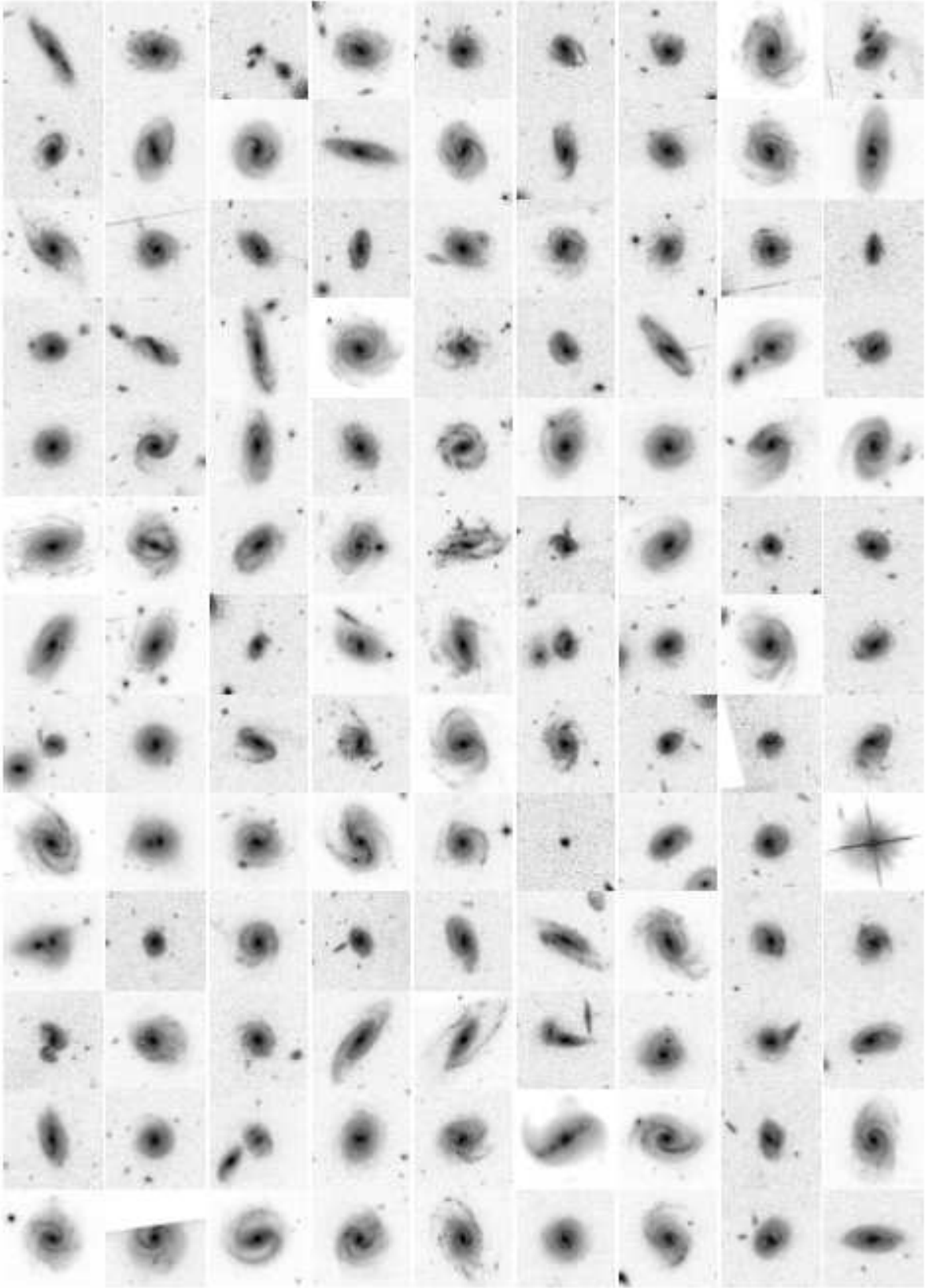}

   \caption{ACS I-band $12''\times 12''$ cut-outs of the non-LAEs with available ACS information, which are the sources in the center of each image. At $z\sim 0.3$, the scale is 4.45 kpc/$''$, which means that the diameter of each cut-out represents 53.4 kpc.
              }
         \label{shapes_control}
   \end{figure*}

Figures \ref{shapes_LAEs} and \ref{shapes_control} show the ACS cut-outs of LAEs and non-LAEs, respectively, with available ACS information. It can be seen that the vast majority of non-LAEs have clear disk-like morphologies, most of them belonging to the Sb, Sc or Sd classes. However, LAEs tend to depart from that tendency. Although two of them seem to be disk-like galaxies, the other six LAEs have compact or interacting/merging morphologies. As  was pointed out before, we only have eight LAEs with available ACS information, so the results are not statistically significant. However, despite the scarcity, the morphology of LAEs seem to be more heterogeneous than that of non-LAEs and this is an indication of a possible morphological differentiation.

\cite{Cowie2011} also studied the sizes and morphology of their LAEs and UV-selected galaxies at $z \sim 0.3$, examining the dependence of size upon the H$\alpha$ equivalent width (EW(H$\alpha$)) as well. They report that LAEs with higher EW(H$\alpha$) are generally unresolved at the 1'' resolution of the CFHT MegaPrime U-band images, while the galaxies with lower EW(H$\alpha$) are generally extended. Furthermore, at the same EW(H$\alpha$), they found that LAEs are significantly smaller than galaxies without Ly$\alpha$ emission. This agrees with the result we find here for our IR-detected LAEs, despite the results were obtained with optical observations in different bands. \cite{Cowie2010} analyzed the morphology of a sample of LAEs and other galaxies without Ly$\alpha$ emission by using i'-band ground-based data from the CFHT MegaPipe database. In agreement with our result, they found that the LAE sample contains a much larger fraction of mergers and compact galaxies then the NUV-continuum selected sample.

These significant morphological differences, along with the results found in previous sections, indicate that Ly$\alpha$ photons tend to escape preferentially from irregular or interacting galaxies of small size, low dust attenuation, and high SFRs.

\section{Conclusions}\label{conclu}

In this study, we have obtained fundamental physical properties (UV and IR emission, dust attenuation, SFR, and morphology) of LAEs at $z\sim 0.3$, defined as those Ly$\alpha$-emitting galaxies with Ly$\alpha$ EW$_{\rm rest-frame}$ above 20 \AA, the typical threshold in the narrow-band searches at higher redshifts. Furthermore, we have compared LAEs with non-LAEs, defined as galaxies at the same redshift as LAEs which, with the same UV and IR limiting luminosities, show no
 Ly$\alpha$ emission with Ly$\alpha$ EW$_{\rm rest-frame}\gtrsim$ 20 \AA\ in their spectrum. The main conclusions of our work are:

\begin{enumerate}

\item We find that a large percentage ($\sim$75\%) of our LAEs have MIPS/PACS IR counterparts under a limiting luminosity of $\log(L_{\rm IR}/L_{\odot})\sim 10.4$. These mid-IR/FIR detections are a direct measurement of their dust emission.

\item We find that 80\% of the LAEs studied at $z\sim 0.3$, both IR-detected and undetected, are normal SF galaxies, $L_{\rm IR}<10^{11}\,L_{\odot}$. We find only six LAEs with 10$^{11}\lesssim L_{\rm IR}[L_{\odot}]\lesssim10^{11.5}$ and none with $L_{\rm IR}>10^{11.6}L_\odot$ or with a ULIRG nature. The finding of a noticeable number of ULIRGs at $z\sim 2.5$ exhibiting Ly$\alpha$ emission suggests that the IR nature of objects selected by means of their Ly$\alpha$ emission changes with redshift. 

\item For each value of $L_{\rm IR}$, LAEs are among the least dusty galaxies at $z\sim 0.3$. The distribution of the dust attenuation in 1200 \AA\, of LAEs and non-LAEs are centered around 1.5 and 2.0, respectively. In this study we have obtained dust attenuation by combining UV and FIR measurements without the uncertainties of rest-frame UV/optically based methods, which do not take into account the dust emission in the FIR.

\item The dust attenuation of objects selected via their Ly$\alpha$ emission is evolving with redshift, from dust-free objects at $z\sim 3.0$ to LAEs with high and low/moderate dust attenuation at $z\sim 2.3$ and $\sim$0.3, respectively. However, it should be noted that the procedures followed in the different studies are not the same and that at $z\gtrsim  3$ very few LAEs have been individually studied without the uncertainties of stacking. The suggested evolution of dust attenuation of LAEs along with the finding that the UV and IR nature of LAEs is changing with redshift, indicates that the physical properties of galaxies selected at different redshifts from their Ly$\alpha$ emission are not the same: either they are evolving or the technique is picking up galaxies of a different nature.

\item LAEs have a much bluer UV continuum slope than non-LAEs, which is compatible with their being among the least dusty objects at their redshift. Furthermore, while most non-LAEs follow the trend of normal SF galaxies, some LAEs seem to be starbursts galaxies, indicating a possible difference in their mode of star formation. The distribution of LAEs between the SF and SB relations indicates that a unique relation between UV continuum slope and dust attenuation does not apply for the whole population. Therefore, the determination of dust attenuation from $\beta$ fails for LAEs at $z \sim 0.3$.

\item The SFR$_{\rm total}$ for LAEs tends to be larger than for non-LAEs. We find a noticeable contribution of the IR emission to the SFR$_{\rm total}$ in LAEs and non-LAEs, about 60\%. Therefore, although LAEs have a bright UV continuum and are among the least dusty galaxies at $z\sim 0.3$, it is essential to take into consideration their dust emission in the FIR for obtaining accurate values of their SFR$_{\rm total}$.

\item The size of LAEs tend to be smaller than those of non-LAEs, with median values of 1.5 and 4.1 kpc, respectively. Despite the low number of LAEs with available ACS information, a visual inspection of their morphologies reveals that they tend to be compact, disk-like, or merging/interacting galaxies (i.e., there is a heterogeneity in morphology) in opposition to non-LAEs, which are mainly disk-like galaxies. These are the most noticeable differences between LAEs and non-LAEs.

\end{enumerate}

\begin{acknowledgements}

We thank the anonymous referee for the comments provided, which have helped us to improve the manuscript. This work was supported by the Spanish Plan Nacional de Astrononom\'ia y Astrof\'isica under grant AYA2008-06311-C02-01. Some of the data presented in this paper were obtained from the Multimission Archive at the Space Telescope Science Institute (MAST). STScI is operated by the Association of Universities for Research in Astronomy, Inc., under NASA contract NAS5-26555. Support for MAST for non-HST data is provided by the NASA Office of Space Science via grant NNX09AF08G and by other grants and contracts. Based on observations made with the European Southern Observatory telescopes obtained from the ESO/ST-ECF Science Archive Facility. {\it Herschel} is an ESA space observatory with science instruments provided by European-led Principal Investigator consortia and with important participation from NASA. The Herschel spacecraft was designed, built, tested, and launched under a contract to ESA managed by the Herschel/Planck Project team by an industrial consortium under the overall responsibility of the prime contractor Thales Alenia Space (Cannes), and including Astrium (Friedrichshafen) responsible for the payload module and for system testing at spacecraft level, Thales Alenia Space (Turin) responsible for the service module, and Astrium (Toulouse) responsible for the telescope, with in excess of a hundred subcontractors. PACS has been developed by a consortium of institutes led by MPE (Germany) and including
UVIE (Austria); KUL, CSL, IMEC (Belgium); CEA, OAMP (France); MPIA (Germany); IFSI, OAP/AOT, OAA/CAISMI,
LENS, SISSA (Italy); IAC (Spain). This development has been supported by the funding agencies BMVIT (Austria), ESA-
PRODEX (Belgium), CEA/CNES (France), DLR (Germany), ASI (Italy), and CICYT/MICINN (Spain). 

\end{acknowledgements}

\end{document}